\documentclass[aps,pra,amsmath,twocolumn,amssymb,floatfix,showpacs,superscriptaddress,nofootinbib,nourl]{revtex4-1} 
\usepackage{tikz}
\usepackage{graphicx}%
\usepackage[left=1.2cm, right=1.2cm, top=1.785cm, bottom=2.0cm]{geometry}
\usepackage{color}
\usepackage{transparent}
\usepackage{psfrag}
\usepackage{dcolumn}
\usepackage{amsmath,wasysym}
\usepackage{amssymb}
\usepackage{latexsym}
\usepackage{booktabs}
\usepackage{latexsym, bbold,bm}
\usepackage[english]{babel}
\usepackage[colorlinks=true,linktoc=page,linkcolor=blue,citecolor=blue,urlcolor=blue]{hyperref}
\usepackage{calrsfs}
\DeclareMathAlphabet{\pazocal}{OMS}{zplm}{m}{n}
\usepackage{tabularx,booktabs}
\usepackage{diagbox}
\newcolumntype{C}{>{\centering\arraybackslash}X}
\begin{document}


\def\bea{\begin{eqnarray}}
\def\eea{\end{eqnarray}}
\def\beq{\begin{equation}}
\def\eeq{\end{equation}}
\def\f{\frac}
\def\k{\kappa}
\def\e{\epsilon}
\def\ve{\varepsilon}
\def\be{\beta}
\def\D{\Delta}
\def\h{\theta}
\def\t{\tau}
\def\a{\alpha}
\def\Pe{{\rm Pe}}

\def\cDa{{\cal D}[X]}
\def\cD{{\cal D}[x]}
\def\cL{{\cal L}}
\def\cLo{{\cal L}_0}
\def\cLa{{\cal L}_1}
\def\rv{{\bf r}}
\def\tv{\hat{\bf{t}}}
\def\on{{\omega_{\rm a}}}
\def\od{{\omega_{\rm d}}}
\def\off{{\omega_{\rm off}}}
\def\fv{{\bf{f}}}
\def\fm{\bf{f}_m}
\def\zh{\hat{z}}
\def\yh{\hat{y}}
\def\xh{\hat{x}}
\def\km{k_{m}}
\def\nh{\hat{n}}

\def\Re{{\rm Re}}
\def\sj{\sum_{j=1}^2}
\def\rk{\rho^{ (k) }}
\def\rek{\rho^{ (1) }}
\def\cek{C^{ (1) }}
\def\rz{\rho^{ (0) }}
\def\rt{\rho^{ (2) }}
\def\rtb{\bar \rho^{ (2) }}
\def\trk{\tilde\rho^{ (k) }}
\def\trek{\tilde\rho^{ (1) }}
\def\trz{\tilde\rho^{ (0) }}
\def\trt{\tilde\rho^{ (2) }}
\def\r{\rho}
\def\tD{\tilde {D}}

\def\s{\sigma}
\def\kb{k_B}
\def\bF{\bar{\cal F}}
\def\F{{\cal F}}
\def\la{\langle}
\def\ra{\rangle}
\def\nn{\nonumber}
\def\up{\uparrow}
\def\dn{\downarrow}
\def\S{\Sigma}
\def\dg{\dagger}
\def\d{\delta}
\def\p{\partial}
\def\l{\lambda}
\def\L{\Lambda}
\def\G{\Gamma}
\def\o{\Omega}
\def\w{\omega}
\def\g{\gamma}
\def\E{{\mathcal E}}

\def\O{\Omega}

\def\vv{ {\bf v}}
\def\jv{ {\bf j}}
\def\jr{ {\bf j}_r}
\def\jd{ {\bf j}_d}
\def\jdd{ { j}_d}
\def\noi{\noindent}
\def\a{\alpha}
\def\d{\delta}
\def\p{\partial} 

\def\la{\langle}
\def\ra{\rangle}
\def\e{\epsilon}
\def\n{\eta}
\def\g{\gamma}
\def\break#1{\pagebreak \vspace*{#1}}
\def\hf{\frac{1}{2}}
\def\rcurs{r_{ij}}

\def\bv{ {\bf b}}
\def\uv{ {\bf u}}
\def\rv{ {\bf r}}
\def\cf{{\mathcal F}}
\def\pg{{\pazocal G}}
\def\pk{{\pazocal K}}
\def\pc{{\pazocal C}}
\def\ord{\psi_N}
\def\kur{\pk_{\ord}}
\def\pn{\pazocal{N}}
\def\uh{{\hat {\bf u}}}
\def\pa{\pazocal{A}}
\def\pb{\pazocal{B}}

\definecolor{darkgreen}{rgb}{0.0, 0.42, 0.24}
\definecolor{royalblue}{rgb}{0.25, 0.41, 0.88}
\definecolor{darkpink}{rgb}{1.0, 0.08, 0.58}
\definecolor{darkcyan}{rgb}{0.0, 1.0, 1.0}
\definecolor{sienna1}{rgb}{1.0, 0.56, 0.0}
\definecolor{grayhtml}{rgb}{0.5, 0.5, 0.5}



\title{Inertia and Activity: Spiral transitions in semi-flexible, self-avoiding polymers}
\author{Chitrak Karan}
\email{chitrak.k@iopb.res.in}
\affiliation{Institute of Physics, Sachivalaya Marg,  Bhubaneswar 751005, India}
\affiliation{Homi Bhabha National Institute, Training School Complex, Anushakti Nagar, Mumbai 400094, India}

\author{Abhishek Chaudhuri}
\email{abhishek@iisermohali.ac.in}
\affiliation{Department of Physical Sciences, Indian Institute of Science Education and Research Mohali, Sector 81, Knowledge City, S. A. S. Nagar, Manauli PO 140306, India}

\author{Debasish Chaudhuri}
\email{debc@iopb.res.in}
\affiliation{Institute of Physics, Sachivalaya Marg,  Bhubaneswar 751005, India}
\affiliation{Homi Bhabha National Institute, Training School Complex, Anushakti Nagar, Mumbai 400094, India}

\date{\today}

\begin{abstract}
We consider a two-dimensional, tangentially active, semi-flexible, self-avoiding polymer to find a dynamical re-entrant transition between motile open chains and spinning achiral spirals with increasing activity. Utilizing probability distributions of the turning number, we ascertain the comparative stability of the spiral structure and present a detailed phase diagram within the activity inertia plane. The onset of spiral formation at low activity levels is governed by a torque balance and is independent of inertia. At higher activities, however, inertial effects lead to spiral destabilization, an effect absent in the overdamped limit. We further delineate alterations in size and shape by analyzing the end-to-end distance distribution and the radius of gyration tensor.The Kullback-Leibler divergence from equilibrium distributions exhibits a non-monotonic relationship with activity, reaching a peak at the most compact spirals characterized by the most persistent spinning. As inertia increases, this divergence from equilibrium diminishes. 
\end{abstract}

\maketitle

\section{Introduction}
\label{sec_intro}

Active polymers are ubiquitous in biological systems~\cite{Alberts2018}. Biofilaments like microtubules and actin filaments constitute the cytoskeleton of the cell, providing its mechanical stability and helping in cell motility and response. On the other hand, the main ingredient of chromosomes in the cell nucleus is the semiflexible polymer DNA. These polymers are driven out of equilibrium by active processes, e.g., treadmilling and active stress generation by molecular motors in crosslinked cytoskeletal filaments or the active drive by RNA polymerase during DNA transcription. Such active processes involve local energy consumption, such as via hydrolysis of ATP or GTP, generating directed motion or stress so that equilibrium fluctuation-dissipation relation and detailed balance are no longer valid. Another prominent example is bacterial locomotion, involving the active rotation of flagella~\cite{Wadhwa2022}. Motivated by such natural examples, a number of reconstituted or artificial active filament systems have been prepared and studied over the last decades~\cite{Sanchez2011,Li2011,Sasaki2014,Vach2015,Martinez2015,Zhang2015}. One prominent example is molecular motor assay, in which cytoskeletal filaments move actively under the influence of conjugate molecular motors attached at one end irreversibly to a substrate, with the other end walking on the filaments~\cite{Schaller2010}. 

In micron-sized molecular systems, the typical inertial time scale $\sim 100$\,ns is much smaller than the typical persistence time $\sim 1$\,s~\cite{Kurzthaler2018}, and thus the inertial effect can be ignored. 
The microscopic dynamics of cilia or flagella~\cite{Bray2000, Fulford1986, Gilpin2020}, actin filaments and microtubules in cytoskeleton~\cite{Fletcher2010, Huber2013a, Prost2015} or artificial motility assays~\cite{Kron1986, Bourdieu1995}, and in Janus colloid chains~\cite{Nishiguchi2018} are describable within overdamped motion of active filaments~~\cite{Sekimoto1995,Laskar2013a, Chelakkot2014, Ghosh2014,Isele2015,Anand2018,Duman2018,Prathyusha2018,Fily2020,Philipps2022}. Cilia-like beating and spiral rotation of overdamped semiflexible filaments at high enough activity were predicted in theoretical studies as observed in earlier motility assay experiments~\cite{Sekimoto1995, Fily2020, Isele2015, Bourdieu1995, Kron1986, Gupta2019, Shee2021, Karan2023}.

Nonetheless, filamentous objects of macroscopic size, including natural examples like millipedes, worms, and snakes and engineered entities such as granular, robotic, or hexbug chains, can exhibit significant inertial effects~\cite{Ozkan-Aydin2021, Garcia2021, Marvi2014, Hu2009, Bhat2023, Safford2009, Zheng2023}. 
This is crucial in shaping these systems' dynamics and steady-state properties. 
 Recent studies suggest that inertia suppresses motility-induced phase separation~\cite{omar23,Mandal2019} and instability in active nematics~\cite{chatterjee2021inertia} and affects diffusivity, mobility, and kinetic temperature in active Brownian particles~\cite{Khali2024, Patel2023, patel2024exact}.  
 However, despite its recognized importance, the role of inertia in active filament motion is not well-understood, with only limited studies highlighting differences between inertial and non-inertial dynamics in flexible chains~\cite{Fazelzadeh2023}. 

In this paper, we carefully study tangentially active semiflexible filaments to investigate the impact of inertia on their conformational states and dynamics.  Using extensive numerical simulations, we obtain a detailed phase diagram as a function of activity and inertia, identifying the two dynamic and conformational phases, e.g., the motile open chain (MOC) and spinning achiral spirals (SAS). We present analytic arguments to describe the observed phase boundaries. In the MOC state, the filament displays persistent motion, with the persistence decreasing inversely with activity. In contrast, the SAS state shows clockwise or counter-clockwise rotation around the filament's center of mass. The rotation speed increases linearly with activity. Our study reveals a remarkable re-entrant transition for inertial active polymers from MOC to SAS to MOC with increasing activity, highlighting the absence of the re-entrance in the overdamped limit. 

The rest of the paper is structured as follows. In section~\ref{sec_model}, we describe the model and the dimensionless control parameters. After that, in section~\ref{sec_results}, we present the numerical analysis and results, scaling arguments, behavior of excess kurtosis, the unwinding process in the model, and size and shapes. Finally, section~\ref{sec_discuss} summarizes the main results with concluding remarks.

\section{Model}
\label{sec_model}

We consider an extensible, semi-flexible, self-avoiding polymer~(Fig.\ref{fig_schematic}$(a)$) of $N$-beads described by monomer positions $\rv_1,~\rv_2,...,~\rv_N$, and bond vectors $\bv_i = \rv_{i+1} - \rv_i$ ($i = 1, 2,...,N-1$) with local tangents $\tv_i = \bv_i/|\bv_i|$. The chain connectivity is maintained by a stretching energy,
\bea
\E_s = \f{k_s}{2} \sum_{i=1}^{N-1} \left[ \bv_i - r_0 \tv_i \right]^2,
\label{eq_ene_stretch}
\eea
where $k_s$ is the spring constant and $r_0$ is the equilibrium bond length. The semi-flexibility is described by the bending rigidity $\k$ and the local bending energy cost
\bea
\E_b = \f{\k}{2 r_0} \sum_{i=1}^{N-2} \left[ \tv_{i+1} - \tv_{i} \right]^2.
\label{eq_ene_bending}
\eea
The self-avoidance is modeled by the short-range Weeks-Chandler-Anderson (WCA) repulsion between the non-bonded pairs of monomers $i$ and $j$,
\bea
\E_{\rm WCA} (r_{ij}) =
\begin{cases}
        4 \e \left[ \left(\f{\s}{r_{ij}}\right)^{12} - \left(\f{\s}{r_{ij}}\right)^{6} + \f{1}{4} \right], & \text{for } r_{ij} < 2^{1/6} \s, \\
        0, & {\rm otherwise},
\end{cases}
\nn\\
\label{eq_energy_wca}
\eea
where $\e$ sets the strength of repulsion and $\s$ the size of repulsive core. 
The total energy cost $\E = \E_s + \E_b + \sum'_{i,j} \E_{\rm WCA} (r_{ij})$ describes the equilibrium semi-flexible polymer, where the last sum is over all non-bonded pair of monomers. 

The activity is implemented via the self-propulsion force ${\bf F}_p$, which is expressed in terms of the constant magnitude $f_a$ along the local tangent, such that
\bea
{\bf F}_p = \sum_{i=1}^{N-1} f_a \tv_i\,.
\label{eq_force_active}
\eea
This implementation is the same as \cite{Sekimoto1995} and differs from \cite{Isele2015} in which the magnitude of the active force depends on the local bond length.

The dynamics of the filament are described by the underdamped Langevin equation for each monomer,
\bea
m \dot \vv_i = -\g(\vv_i - v_0 \uv_i) - \bm{\nabla}_i \E + \g \sqrt{2D}~\bm{\eta}_i(t),
\label{eq_langevin}
\eea
where $\vv_i$ is the velocity of $i$-th particle, and $\uv_i = (\tv_{i-1} + \tv_i)/2$ with $i=2, \dots, N-1$, with the boundary terms $\uv_1=\tv_1/2$, $\uv_N=\tv_{N-1}/2$. Note that the tangential self-propulsion acting along the bonds leads to this effective active force on individual monomers. 
Moreover, the equilibrium diffusion constant $D=k_B T/\g$, and the Gaussian random noise obeys $\la \bm{\eta}_i(t) \ra=0$ and $\la \bm{\eta}_i(t) \otimes \bm{\eta}_j(0) \ra = \d_{ij} \d(t) \mathbb{1}$. In the above equation, we used $f_a=\g v_0$ with $v_0$ denoting the self-propulsion speed. 
The length, energy, and time scales are set by $\s$, $\kb T$, and $\t_d = \g \s^2 / \kb T$, the time to diffuse over $\s$. The dynamics is controlled by the microscopic inertial relaxation time $\t_m = m/\g$, and active time $\t_a = \s / v_0$, which can be expressed in terms of the following two dimensionless control parameters  
\bea
M = \f{\t_m}{\t_d} = \f{m \kb T}{\g^2 \s^2},~{\rm and}
\label{eq_M}
\eea
\bea
\Pe = \left( \f{\t_a}{\t_d} \right)^{-1} = \f{f_a \s}{\kb T}.
\label{eq_Pe0}
\eea
We perform numerical simulations of $N=64$ bead chains in two dimensions~(2d) using the velocity-Verlet scheme. We use $\e = \kb T$ and $r_0 = \s$ for simplicity. The choice of spring constant $k_s = 10^3 \kb T /r_0^2$, keeps the relative mean-squared fluctuation of bond lengths $\sqrt{\la \d b^2 \ra}/r_0 < 4\%$ for a chain of mean length $L=(N-1)r_0$, where $\d b$ is the deviation of bond length from $r_0$. 
For the worm-like chain model of semiflexible polymers, the equilibrium persistence length $\ell_p=2\k/k_B T$ in two dimensions. The persistence parameter $u=L/\ell_p$ controls if the chain behaves like a rigid rod ($u\approx 1$) or flexible chain ($u\approx 10$). In the intermediate parameter values, the effect of semiflexibility is most pronounced, e.g., in 2d, the regime $3 \lesssim u \lesssim 4$ displays clear double minima in the Helmholtz free energy corresponding to the coexistence of rigid-rod and flexible chain behaviors in the equilibrium~\cite{Dhar2002, Chaudhuri2007}. We choose $u = 3.33$ in this study.

We express Eq.\eqref{eq_langevin} in the following dimensionless form, to perform the numerical simulations
\bea
M \f{d \tilde \vv_i}{d \tilde t} = - ( \tilde \vv_i - \Pe\, \uv_i ) - \tilde{\bm{\nabla}}_i U + \tilde{\bm{\eta}}_i(t),
\label{eq_langevin_dimless}
\eea
where $\tilde{\vv}_i = \vv_i \t_d/\s$, $\tilde{t} = t/\t_d$, $\tilde{\bm{\nabla}}_i = \s \bm{\nabla}_i$, $U = \E/(\kb T)$, and $\tilde{\bm{\eta}}_i = \sqrt{\t_d} \bm{\eta}_i$.
In the integration, we use a time-step size $\d {\tilde t} = 10^{-4}$ in all the simulations. The results presented in this paper used simulations over $10^{10}$ time steps with gaps of $10^4$ steps to obtain uncorrelated statistics.

\section{Results}
\label{sec_results}

\begin{figure*}[t!]
\includegraphics[width=2\columnwidth,height=!]{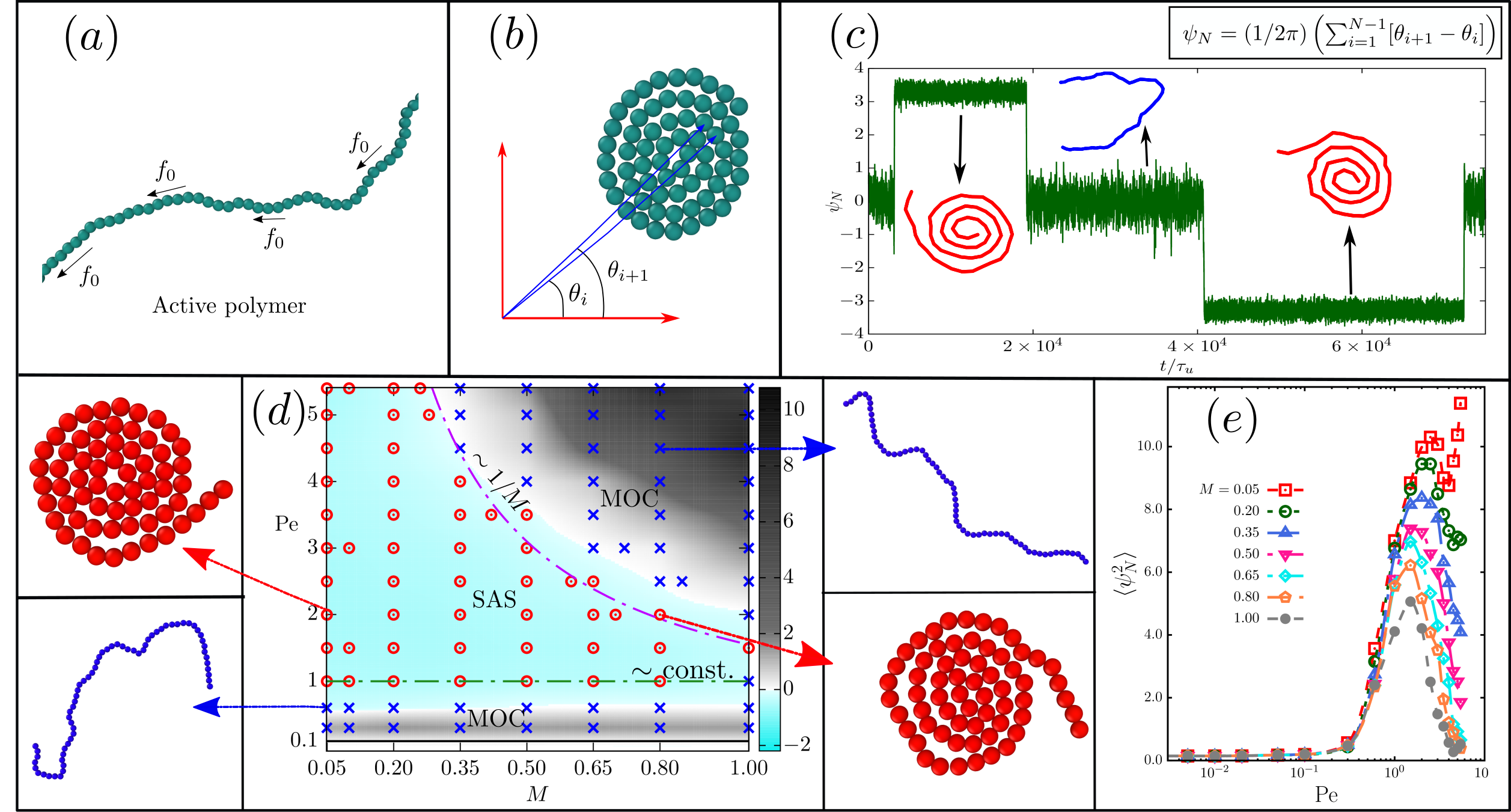}
\caption{(colour onlie) $(a)$ Schematic diagram of tangentially driven polymer in two dimensions. $(b)$ Calculation of the differential turning number of the chain between two consecutive monomers. $(c)$ Typical time series of the turning number with different steady-state representations. $(d)$ $M$ vs. $\Pe$ phase diagram. Cross~({\color{blue} \pmb{$\times$}}) denotes stable open and meta-stable spiral phase. Circle-dot~({\color{red} \pmb{$\odot$}}) denotes stable spiral and meta-stable open phase. Background colors denote the excess kurtosis ($\kur$) with a color bar showing the numerical values. Configurations of stable states of corresponding points are shown. In the phase diagram, the green dashed line approximately denotes the transition from the meta-stable spiral to the stable spiral phase, and the pink dashed denotes the transition from the stable spiral to the meta-stable spiral phase, showing the scaling obtained in Eq.\eqref{eq_Pe2}. Black solid line at $\Pe=0.1$ denotes the $\Pe_c$ obtained in Eq.\eqref{eq_Pe_onset}. $(e)$ The variance $\la \ord^2 \ra$ as a function of $\Pe$ at different values of $M$ denoted in the figure legend. The non-monotonicity at higher $M$ indicates re-entrant transition.}
\label{fig_schematic}
\end{figure*}

In the $\Pe$-$M$ phase space, we observe spontaneous spiral formation in a wide range of parameter values.  The spirals characterized by the turning number~(Fig.~\ref{fig_schematic}$(b)$) $\ord = (1/2\pi) \left( \sum_{i=1}^{N-1} [ \theta_{i+1} -\theta_i ] \right)$, where $\theta_i$ is the angle subtended by the local tangent $\tv_i = \cos{\theta_i} {\hat x} + \sin{\theta_i} {\hat y}$. The quantity $\ord$ calculates the number of turns the chain takes between two ends, with $\ord = 0$ denoting an open chain configuration. The negative and positive values of $\ord$ denote clockwise and counter-clockwise turns in the spiral.

The polymer dynamically transits between spiral and open states at adequately high activity. Fig.~\ref{fig_schematic}$(c)$ shows a typical time series of $\ord$ where three possible states appear: an open state where $\ord$ fluctuates around zero, and clockwise and anti-clockwise spiral states with equal and opposite values of $\ord$. 
In the absence of chirality, both kinds of spirals are equally likely. 

One can calculate the steady-state distributions of $\ord$ from the time series at different $\Pe$ and $M$ values. For the achiral system under consideration, this distribution is symmetric around $\ord=0$ and, in general, shows three maxima, one at $\ord= 0$, denoting the open state,  and the other two maxima of equal heights at $\ord=\pm \l$, denoting the clockwise and counter-clockwise spiral states. 
We identify two distinct phases from the relative weights: $(a)$ motile open chain (MOC) in which the open state is more probable, and $(b)$ spinning achiral spirals (SAS) in which the spiral state is more probable.

\subsection{$\Pe-M$ phase diagram}
\label{sec_phdia1}

Fig.~\ref{fig_schematic}$(d)$ shows the phase diagram for the active polymer with the variation of control parameters $\Pe$ and $M$.  The phase diagram shows regions of two phases, SAS and MOC, having two boundaries: 
$(a)$~The first boundary at lower $\Pe$ is independent of $M$. This line characterizes the transition from a stable MOC to a stable SAS.
($b$)~The second boundary at higher $\Pe$ which depends on $M$ as $1/M$ marks a re-entrant transition from stable SAS to stable MOC. 
   
The phase diagram also shows a heat map of the excess kurtosis of turning number $\kur$, defined as 
\bea
\kur = \f{\la \ord^4 \ra}{3 \la \ord^2 \ra^2} - 1.
\label{eq_kurt}
\eea
As we show later, the vanishing and positive $\kur$ regimes correspond to MOC, and the negative values denote SAS. A detailed analysis of kurtosis and its relevance in accurately identifying the phases is presented in Sec.\ref{sec_kurt}. 

The first phase boundary describing the onset of spiral formation can be estimated using the torque balance condition. For a SAS of radius $R$, it reads $\Gamma \w = f_a R - \k/R$ interpreting $f_a$ as the self-propulsion force acting on each monomer, and using $\G$ to represent the rotational drag and $\w$ denoting rotational speed of the spiral. At the onset of SAS, $\w=0$, we obtain the condition 
\bea
f_a^{(c)} = \k/R^2.
\label{eq_fc}
\eea
This corresponds to a critical P{\'e}clet, $\Pe_c= f_a^{(c)}\s/\kb T$, which determines the onset of spirals for $\Pe > \Pe_c$. This value depends on the bending rigidity $\k$ but is independent of $M$. Such a prediction agrees with the small $\Pe$ phase boundary in Fig.~\ref{fig_schematic}$(d)$.

For larger $f_a$ the spiral will start to rotate. At the beginning of the rotation, the viscous drag $\G\w =  \tilde R (f_a - f_a^{(c)})$ 
leading to 
\bea
\w \approx  \f{\tilde R}{\G} 
(f_a - f_a^{(c)}),
\label{eq_omega}
\eea
a behavior corroborated in Fig.~\ref{fig_tangent_corr}($c$).
For a single turn of spiral $\tilde R=L/2\pi$. 
Therefore, at the onset of the spiral formation
\beq
\Pe_c = \f{f_a^{(c)} \s}{k_B T} = \f{\ell_p \s}{2 \tilde R^2} = \f{2\pi^2 \s}{u L}
\label{eq_Pe_onset}
\eeq
with $u=L/\ell_p$. For the choice of $u=3.33$ and $L=(N-1)\s$ with $N=64$, we get $\Pe_c \approx 0.1$. This estimate agrees well with the fitting of Fig.\ref{fig_tangent_corr}($c$). This is the minimum $\Pe$ required for spiral formation, but spirals become a dominant state at a far higher $\Pe \sim 1$, as indicated in the phase diagram.

Increasing $\Pe$ generates higher torque, leading to more compact spirals. Such compaction reduces available space at the center of a spiral, reducing the probability of the leading tip turning stochastically in the wrong direction and opening up. As a result, the stability of the spirals increases as we increase $\Pe$ from small to intermediate values; see Fig.~\ref{fig_schematic}$(e)$. The absolute turning number $\la |\ord| \ra$ of the spiral increases with $\Pe$, and the spiral gets tighter~(see Appendix-\ref{app_spiral}). However, a further higher $\Pe$ destabilizes the spirals, reducing the probability of their appearance and leading to a re-entrant transition to MOC at a higher $\Pe$. The mean turning number $\la |\ord| \ra$ of the spirals decreases with $M$~(see Appendix-\ref{app_spiral}). We must carefully consider their dynamic unwinding mechanism to understand the spiral instability. 

\begin{figure}[t] 
\includegraphics[width=\columnwidth,height=!]{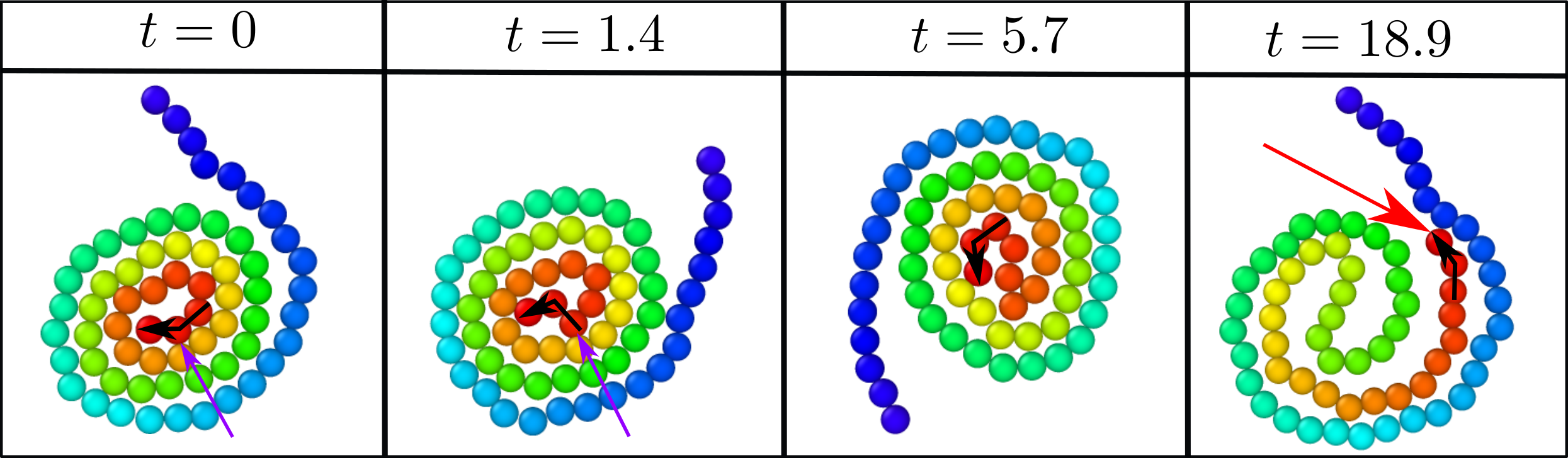}
\caption{(colour online) Unwinding dynamics for $M=1$ and $\Pe=3$. The chain's color gradient denotes the chain's monomer index, with red as the head and blue as the tail. Black arrowheads show the propulsion directions of the polymer head. Violet arrows indicate the particle's direction of motion that triggers the unwinding process. The red arrow in the last panel points out the head.}
\label{fig_unwind}
\end{figure}

Fig.~\ref{fig_unwind} shows snapshots of the spiral unwinding process. The clockwise rotating leading tip, denoted by the black arrow, turns around spontaneously in the anti-clockwise direction, destabilizing the spiral. This eventually unwinds the whole spiral. For the leading tip turnaround, it is necessary to have enough open space at the center of the spiral. In the presence of inertia, high activity leads to enhanced inertial recoil, allowing the self-propelling tip to move away from its propulsion direction, creating more space and opportunities for the turnarounds that destabilize the spiral.

To understand this behavior, we must consider the effect of inertial recoil on the spiral. 
As we increase activity, $\t_a$ decreases, increasing the collision frequency. Collisions of the leading tip with the nearest segments impart force. This picture is similar to a ball bouncing on a wall in the presence of gravitational force. Within this setting, the equation of motion of the leading tip having mass $m$ can be expressed as
\bea
m \f{dv}{dt} = -\g v + f_a - f_a \t_c \d(t-\t_c),
\label{eq_eom}
\eea
with collision time $\t_c \approx \ell/v_0 \propto \t_a$ for a gap $\ell$, and assuming that the opposite impulse at collision is given by the active force $f_a$ due to the surrounding particles.  A straightforward calculation shows,
\bea
v(t) = \f{f_a}{\g} - \f{f_a \t_c}{m} \exp\left(- (t-\t_c)/\t_m \right),
\eea
where the inertial relaxation time $\t_m=m/\g$.
The recoil allows the velocity to change sign at collision, $t=\t_c$, to give  
$f_a(1/\g - \ell/v_0 m)=0$. Using $\ell=\s$, this leads to the condition $mv_0/\g\s=1$, i.e., $\t_m/\t_a=1$.

As long as $\t_a > \t_m$, inertial recoil does not affect the dynamics, and active motion is regained between consecutive collisions.  In the presence of finite inertia, as we increase $\Pe$, $\t_a$ decreases, and the probability of turnarounds due to recoil increases. As a result, SAS state gets destabilized as $\t_a \lesssim \t_m$. This leads to the estimate of the re-entrant phase boundary,  
\bea
\Pe \propto 1/M.
\label{eq_Pe2}
\eea
This prediction shows reasonable agreement with the upper phase boundary in Fig.~\ref{fig_schematic}$(d)$. 

On the other hand, in the overdamped limit of $\t_m=0$, the terminal speed $v_0=f_a/\g$ is regained instantaneously after every collision. In the absence of recoil, this does not allow for the change in orientation and thereby suppresses any possibility for re-entrance.

\subsubsection{Dynamical characterization: center of mass motion}

\begin{figure}[t!]
\includegraphics[width=\columnwidth,height=!]{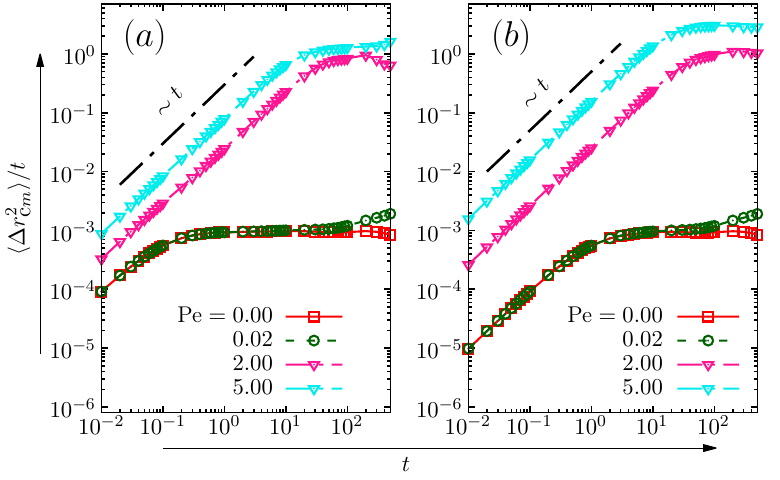}
\caption{(colour onlie) Mean-squared displacement for $(a)$ $M=0.05$ and $(b)$ $M=0.5$ at different $\Pe$-values mentioned in the plots. Black dashed lines plot guide to eye for $\sim t$ scaling.}
\label{fig_msd}
\end{figure}

Here, we use the mean-squared displacement (MSD) of the filament center of mass (COM) to characterize the dynamical properties of the chain in various phases~(Fig.\ref{fig_msd}). The initial evolution of all the MSD at non-zero $\Pe$ shows ballistic-diffusive-ballistic crossovers. The first ballistic-diffusive crossover is determined by the inertial relaxation time $M$. Thus, it shifts to a later time for higher inertia; see Fig.\ref{fig_msd}($b$).  Asymptotically, at time scales beyond all correlation times, one expects effective diffusive behavior~\cite{Gupta2019}. The details of such crossovers can involve the relaxation of an emergent COM persistence and fluctuations in effective COM speed~\cite{Gupta2019}. For the current purpose, it suffices to note that the second ballistic regime is due to the active drive and is absent in the equilibrium chain. It is dominated by the persistent transport of the pure or coexistent open-chain state.  It's value increases with $\Pe$. At $\Pe \sim 1$, where the dominant state is spiral, the COM of spirals undergoes effective diffusion. However, the overall MSD, even at this phase, shows the second ballistic regime due to the coexisting open chains. Thus, the open chain state in this study is characterized by the motile open chain (MOC).

\subsubsection{Dynamical characterization:  active rotation of spiral}

\begin{figure}[t!]
\includegraphics[width=\columnwidth,height=!]{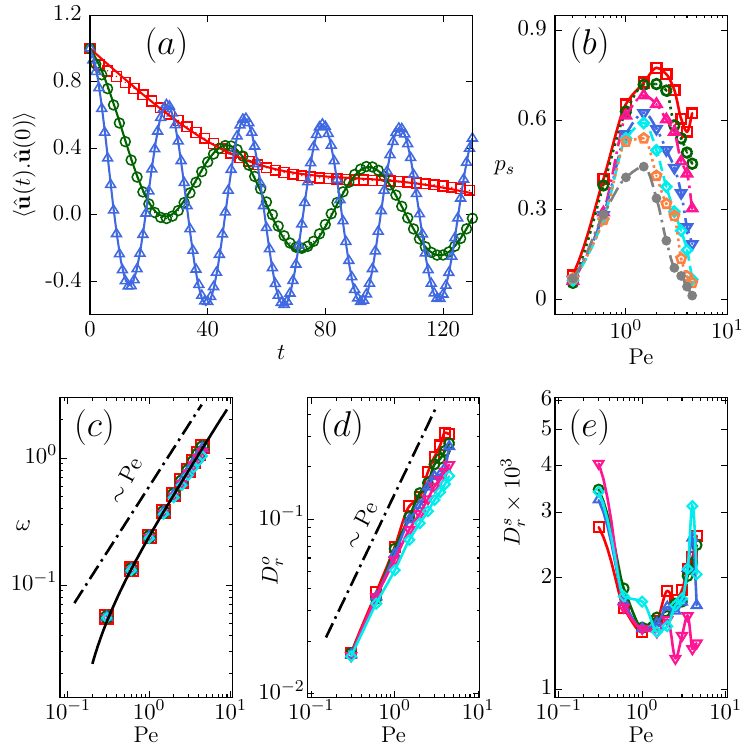}
\caption{(colour onlie) $(a)$ Two-time correlation of mean unit tangent vector, $\la \uh(t) . \uh(0) \ra$. Points denote the numerically calculated results for $M=0.50$ at different $\Pe$ values, $\Pe=0.3$ ({\color{red} \pmb{$\boxdot$}}), $\Pe=0.6$ ({\color{darkgreen} \pmb{$\odot$}}), and $\Pe=1$ ({\color{royalblue} \pmb{$\bigtriangleup$}}). Corresponding solid lines plot the fitting of numerical results with $C_{\uh}(t)$ in Eq.\eqref{eq_corr_fitting}. 
$(b)$ The probability of spiral state $p_s$ as a function of $\Pe$ at different $M$ values,  $M=0.05$~({\color{red} \pmb{$\boxdot$}}), $M=0.20$~({\color{darkgreen} \pmb{$\odot$}}), $M=0.35$~({\color{darkpink} \pmb{$\bigtriangleup$}}), $M=0.50$~({\color{royalblue} \pmb{$\bigtriangledown$}}), $M=0.65$~({\color{darkcyan} \pmb{$\diamondsuit$}}), $M=0.80$~({\color{sienna1} \pmb{$\pentagon$}}), and  $M=1.00$~({\color{grayhtml} \pmb{$\bullet$}}). $p_s$ varies non-monotonically with $\Pe$ and decreases with $M$. 
$(c)$ Rotational speed of spirals as function of $\Pe$ for different $M$ values, $M=0.05$ ({\color{red} \pmb{$\boxdot$}}), $M=0.35$ ({\color{darkgreen} \pmb{$\odot$}}), $M=0.50$ ({\color{royalblue} \pmb{$\bigtriangleup$}}), $M=0.65$ ({\color{darkpink} \pmb{$\bigtriangledown$}}), $M=1.00$ ({\color{darkcyan} \pmb{$\diamondsuit$}}). The black solid line plots $\w = \pc (\Pe-\Pe_c)$ treating $\pc$ and $\Pe_c$ as fitting parameters. 
$(d),~(e)$ Effective rotational diffusivities, $D_r^o$ and $D_r^s$ in Eq.\eqref{eq_corr_fitting}, respectively, as functions of $\Pe$ for different $M$; i.e., $M=0.05$ ({\color{red} \pmb{$\boxdot$}}), $M=0.20$ ({\color{darkgreen} \pmb{$\odot$}}), $M=0.35$ ({\color{royalblue} \pmb{$\bigtriangleup$}}), $M=0.65$ ({\color{darkpink} \pmb{$\bigtriangledown$}}), $M=1.00$ ({\color{darkcyan} \pmb{$\diamondsuit$}}). Lines in these two plots are guides to the eye. }
\label{fig_tangent_corr}
\end{figure}

In the spiral state, the chain spins either clockwise or counter-clockwise depending on the sign of $\psi_N$. The spinning spiral's rotational velocity $\w$ can be directly extracted using the dynamics of the local tangent $\tv_i(t)$.
We employ the two-time correlation $\la \uh(t) \cdot \uh(0) \ra := \f{1}{N-1} \sum_{i=1}^{N-1} \la \tv_i(t) \cdot \tv_i(0)\ra$. While the chain spins in the spiral state, this correlation is expected to show sinusoidal oscillations with an amplitude that decays due to an eventual loss of persistence. Moreover, as the chain dynamically transits between the open and spiral states, we also expect a contribution from the open state in which the correlation is expected to decay exponentially with a persistence time. Using a probability for the spiral state $p_s$ and of open state $1-p_s$, we use the following expression 
\bea
C_{\uh}(t) = (1 - p_s) \exp(- D_r^{o} t) + p_s \exp(- D_r^{s} t) \cos(\w t), \nn\\
\label{eq_corr_fitting}
\eea
for the correlation and fit it to simulation results, to extract the inverse persistence times  $D_r^{o}$ and $D_r^{s}$, along with $p_s$ and the rotational speed $\w$; see Fig.~\ref{fig_tangent_corr}($a$). As Fig.~\ref{fig_tangent_corr}($b$) shows, the probability of spiral state $p_s$ varies non-monotonically with $\Pe$ and decreases with $M$. The rotational velocity increases linearly with $\Pe$ at high activity; see Fig.~\ref{fig_tangent_corr}($c$). Moreover, the data shows good fit to $\w = \pc (\Pe-\Pe_c)$, a form same as Eq.\eqref{eq_omega}, with the proportionality constant $\pc= \f{\tilde R}{\G}= 0.27$ and the spiral onset boundary $\Pe_c=0.11$. This agrees well with the analytic estimate of $\Pe_c \approx 0.1$ obtained in Sec.\ref{sec_phdia1}. Increasing activity renders increased directional noise in the open state. As a result, $D_r^o$, the inverse persistence time in the open state increases as $\sim \Pe$~(Fig.~\ref{fig_tangent_corr}($d$)\,). On the other hand, it varies non-monotonically with $\Pe$ in the spiral state~(Fig.~\ref{fig_tangent_corr}($e$)\,). The initial increase in spiral persistence ($1/D^s_r$) is associated with its initial compaction. The SAS state is most stable with the highest spinning persistence near $\Pe\approx 1$. The subsequent decrease in persistence  ($1/D^s_r$) at higher activity is due to the inertial recoil that destabilizes the spiral.

\subsection{Excess kurtosis}
\label{sec_kurt}

\begin{figure*}[t!]
\centering
\includegraphics[width=2\columnwidth,height=!]{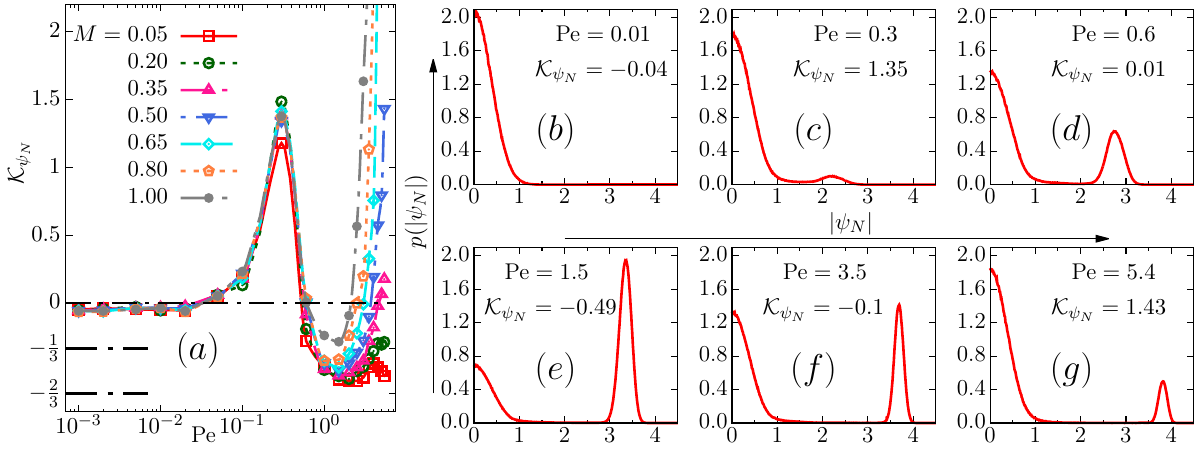}
\caption{(colour onlie) $(a)$ Excess kurtosis as a function of $\Pe$ for different $M$ mentioned in the legend. Dashed line plots $\kur = 0$, which is true for a typical Gaussian system. Lower solid lines plot $\pk_{\ord} = -2/3$ and $\pk_{\ord} = -1/3$. 
$(b)-(g)$ represent probability distributions of the turning number, $p(|\ord|)$, for $M=0.5$ at different $\Pe$-values mentioned in the plots. The associated $\kur$ values are also mentioned.}
\label{fig_spiral_dist}
\end{figure*}

Irrespective of the state, due to chiral symmetry in the system, the mean turning number $\la \ord \ra=0$. 
To quantify the departure of  $\ord$ distributions from the Gaussian nature, we calculate the excess kurtosis of this quantity with mean zero as defined in Eq.\eqref{eq_kurt}. 
The value $\kur=0$ characterizes a Gaussian distribution. In the case of unimodal distributions, positive $\kur$ corresponds to long-tailed distributions with tails asymptotically approaching zero more slowly than Gaussian, and negative $\kur$ suggests a finite support for the distribution, with less extreme outliers than Gaussian. This simple interpretation does not apply in the presence of multimodality. The results of our model show two qualitatively distinct possibilities -- open-chain conformations with unimodal distributions $p(\ord)$ with the maximum at $\ord = 0$, or coexistence of this behavior with distinctly non-equilibrium spiral conformations with equal probabilities for clockwise and anti-clockwise spirals. This second, more generic state shows three maxima, one at $\ord = 0$ and two others at $\ord=\pm \l$. For different $\Pe$ and $M$, the relative probabilities of spiral and open states change. 

To gain an insight into the observed properties of $\kur$, let us consider a linear combination of three Gaussian distributions having maxima at $0$ and $\pm \l$ to approximate the trimodal distributions of $\ord$ at coexistence,
\bea
p_3(\ord) = \f{\pa}{2}~\left[ \pg_{-\l,\s_s}(\ord) + \pg_{\l,\s_s}(\ord) \right] + ( 1 - \pa )~\pg_{0,\s_o}(\ord), \nn\\
\label{eq_triple_gauss}
\eea
with, $\pg_{\l, \s}(\ord)$ denoting Gaussian distribution with mean $\l$ and variance $\s$. Here, we use $\pa$ and $(1-\pa)$ to denote probabilities of the spiral and open state, respectively. Thus, within this theoretical approximation, we expect $\pa = p_s$, with $p_s$ as used in Eq.\eqref{eq_corr_fitting}. We still use two different symbols for these two quantities as they are only theoretical approximations of a more complex phenomenon. However, as we find later~(Fig.\ref{fig_app2} in Sec.\ref{app_prob}), the numerical values of $\pa$ and $p_s$ closely follow each other, reinforcing confidence in the approximate theoretical picture developed here.  

Using the properties of Gaussian distribution, we get
\bea
\la \ord^2 \ra &=& \pa~\left(\s_s^2 + \l^2 \right) + (1 - \pa) \s_o^2,
\nn\\
\la \ord^4 \ra &=& \pa \left( 3 \s_s^4 + 6 \s_s^2 \l^2 + \l^4 \right) +  (1 - \pa)\, 3\s_o^4 .
\label{eq_x2_x4_triple_gauss}
\eea
Substituting these expressions in Eq.\eqref{eq_kurt}, we obtain
\bea
\kur = \f{\pa \left( 3 \s_s^4 + 6 \s_s^2 \l^2 + \l^4 \right) +  (1 - \pa)\, 3\s_o^2}{3[\pa~\left(\s_s^2 + \l^2 \right) + (1 - \pa) \s_o^2]^2} -1.  \nn\\
\label{eq_kurt_triple_gauss}
\eea
For the pure open chain state with the single central Gaussian, $\pa = 0$,  we get $\kur=0$. In the other limit of pure spiral state, $\pa = 1$, we obtain the following simple form $\kur=-\f{2\l^4}{3(\l^2+\s_s^2)^2}$. 

In the Dirac-delta function limit of the distribution of spiral state, $\s_s \to 0$, this expression reduces to $\kur=-\f{2}{3}$. 
When both the spiral and open states are equally probable, setting $\pa=1/2$ and $\s_o=\s_s$ we get $\kur=-\f{\l^4}{3(\l^2+2\s_s^2)^2}$. This expression, in the simplifying limit of three Dirac-delta functions with $\s_s \to 0$, leads to the estimate $\kur=-\f{1}{3}$, an intermediate value lying between the purely open chain value $\kur=0$ and purely spiral state value $\kur=-\f{2}{3}$. These three limiting values set useful points of comparison and are indicated in Fig.\ref{fig_spiral_dist}($a$).
 
We study the excess kurtosis of the turning number in more detail in Fig.~\ref{fig_spiral_dist}($a$). It shows non-monotonic variations of excess kurtosis with increasing $\Pe$ at different inertia. The corresponding variation of the probability distribution of turning number amplitude $p(|\ord|)$ with increasing $\Pe$ is also shown at a representative inertia value $M=0.5$; see Fig.~\ref{fig_spiral_dist}($b$)--$(g)$. For small $\Pe$, the excess kurtosis is vanishingly small, and the probability distribution remains approximately Gaussian with the mode at $|\ord|=0$, corresponding to an open chain. As $\Pe$ increases, a small local maximum in the distribution arises at non-zero $|\ord|$ corresponding to spiral, increasing the excess kurtosis to a positive value. This behavior is approximately similar to the appearance of a non-Gaussian longer tail in an unimodal distribution. The excess kurtosis value starts to decrease as the spiral maximum gets more prominent. It eventually becomes negative and reaches a minimum. Note that the minimum value stays between $\kur=-\f{1}{3}$ and $\kur=-\f{2}{3}$, corresponding to the idealized Dirac delta-function limits of coexistence, having equal probability of the two states, and pure spiral. Beyond this point, with increasing $\Pe$, the spiral state gets destabilized, the non-zero $|\ord|$ maximum starts to decrease, and $\kur$ starts to increase again. This sets the onset of re-entrance.  Further, note that the excess kurtosis of the polymers with different inertia values are not distinguishable at low $\Pe<0.3$. They diverge from each other at higher $\Pe$. Similar emergence of different physical properties in other systems with activity and inertia have been observed in other recent studies~\cite{Mandal2019,Fazelzadeh2023}. 

Further, we fit $p_3(\ord)$ in Eq.\eqref{eq_triple_gauss} to numerically calculated distribution functions to obtain the probability of spirals $\pa$ and thereby obtain estimates of the excess kurtosis using Eq.\eqref{eq_kurt_triple_gauss}. The following tables provide (Table-\ref{tab_kurt}) a comparison between the excess kurtosis obtained using this method and from direct measurements from simulations, and (Table-\ref{tab_a_b}) lists the values of $\pa$. Note that an independent estimate of spiral probabilities $p_s$ can be found from Eq.\eqref{eq_corr_fitting}. This estimate closely follows the dependence of $\pa$ on $\Pe$~(Fig.\ref{fig_app2} in Appendix-\ref{app_prob}), suggesting that they correspond to indeed the same measure. 

\begin{table}[hbtp] 
\begin{tabular}{|l||*{5}{c|}}\hline
\backslashbox{$M$}{$\Pe$}
&\makebox[5.7em]{0.3}&\makebox[5.7em]{0.6}&\makebox[5.7em]{1.5}&\makebox[5.7em]{4.5}\\\hline\hline
0.05 & 1.21~(1.18) & -0.28~(-0.28) & -0.56~(-0.56) &  -0.47~(-0.47) \\\hline
0.50 & 1.42~(1.35) & 0.00~(0.00) & -0.50~(-0.49) & 0.56~(0.57) \\\hline
1.00 & 1.45~(1.38) & 0.00~(0.00) & -0.30~(-0.28) & 7.34~(7.61) \\\hline
\end{tabular}
\caption{A table of excess kurtosis values: numbers outside brackets are estimated from the fitting of simulated distributions to Eq.\eqref{eq_triple_gauss}. The directly measured excess kurtosis values are shown in the parenthesis for comparison.}
\label{tab_kurt}
\end{table}

\begin{table}[hbtp] 
\begin{tabular}{|l||*{5}{c|}}\hline
\backslashbox{$M$}{$\Pe$}
&\makebox[5.7em]{0.3}&\makebox[5.7em]{0.6}&\makebox[5.7em]{1.5}&\makebox[5.7em]{4.5}\\\hline\hline
0.05 & $\pa = 0.10$ & $\pa = 0.45$ & $\pa = 0.76$ &  $\pa = 0.62$ \\\hline
0.50 & $\pa = 0.08$ & $\pa = 0.31$ & $\pa = 0.66$ & $\pa = 0.20$ \\\hline
1.00 & $\pa = 0.08$ & $\pa = 0.31$ & $\pa = 0.46$ & $\pa = 0.01$ \\\hline
\end{tabular}
\caption{Probabilities of spirals $\pa$ extracted from the fitting of Eq.\eqref{eq_triple_gauss} with direct numerical measurement.}
\label{tab_a_b}
\end{table}

\subsection{Size and shape}
\label{sec_size_shape}

Associated with the open-to-spiral transitions, in this section, we characterize the size and shape of the polymer with the help of $(a)$~the end-to-end separation and $(b)$~the radius of the gyration tensor.

\subsubsection{End-to-end separation}

\begin{figure}[t!]
\includegraphics[width=\columnwidth,height=!]{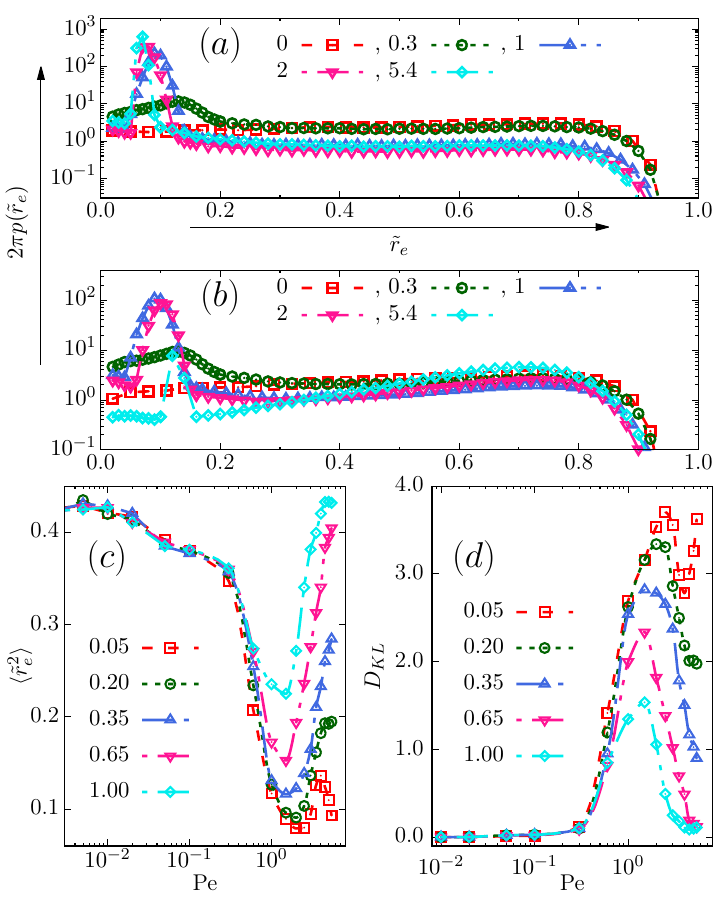}
\caption{(colour onlie) Scaled end-to-end distance, $\tilde r_e = r_e/L$. $(a),(b)$ are distributions of $\tilde r_e$ for for $M=0.05$ and $M=1.00$, respectively. Legends in the plots define the $\Pe$ values. $(c),(d)$ Mean-squared end-to-end distance, $\la \tilde r_e^2 \ra$, and, Kullback-Leibler divergence, $D_{KL}$, as function of $\Pe$ for different $M$ values mentioned in the legend.} 
\label{fig_Ree}
\end{figure}

In Fig.~\ref{fig_Ree}$(a),~(b)$ we show the distributions of scaled end-to-end separation, $\tilde r_e = r_e/L$ where $r_e = |{\rv}(L) - \rv(0)|$, for different $\Pe$ at small and large inertia values $M=0.05$ and $M=1$ respectively. The end-to-end distribution function $p({\tilde r_e})$ is normalized as $\int_0^1 d {\tilde r_e}~2 \pi {\tilde r_e}\, p({\tilde r_e}) = 1$. At equilibrium $\Pe=0$, the distribution shows a shallow maximum near $\tilde r_e=0.8$ corresponding to the open chain due to the semiflexibility of the chain. A second maximum at $\tilde r_e=0$ possible for the ideal worm-like chain~\cite{Dhar2002, Chaudhuri2007} is suppressed due to the self-avoidance of the polymer we considered.  As we increase $\Pe$, a new and more pronounced maximum appears near $\tilde r_e \lesssim 0.1$. This maximum corresponds to a coexistence of the open chain with the SAS state. In the overdamped limit of $M=0.05$, the position of this maximum shifts toward smaller $\tilde r_e$ values, and the peak height increases as the spiral shrinks and gets tighter and more stable with increasing $\Pe$.  In contrast, for larger inertia  $M=1$, the location and height of this second maximum near $\tilde r_e \approx 0.1$ varies non-monotonically with increasing $\Pe$, associated with the re-entrant transition.  At small $\Pe$, the spiral gets tighter as in the overdamped case; however, beyond $\Pe \simeq 1 $, the spiral peak position shifts to higher $\tilde r_e$, and the peak height decreases as the spirals start to open up. 

We also show in Fig.~\ref{fig_Ree}$(c)$ the non-monotonic variations of $\la \tilde r_e^2 \ra$ with $\Pe$ for different $M$ values. At first, with increasing activity, the open chain gets destabilized towards the spiral state, and $\la \tilde{r}^2_e\ra$ reduces until $\Pe \sim 1$. Beyond that, the inertial chain shows swelling and consequent onset of instability of the spiral structure towards the open chain with increasing $\Pe$, and $\la \tilde r_e^2 \ra$ increases. This second instability disappears in the overdamped limit of $M=0.05$, and $\la \tilde r_e^2 \ra$ saturates. With increasing $M$, the compaction of $\la \tilde r_e^2 \ra$ deep inside the spiral state near $\Pe \sim1 $ gets less pronounced.

In Fig.~\ref{fig_Ree}$(d)$, we present the Kullback-Leibler divergence from equilibrium using the end-to-end distribution functions by calculating,
\bea
D_{KL} = \int_0^1 d{\tilde r_e}~p({\tilde r_e})~\ln\left[ \f{p({\tilde r_e})}{p_{\rm eq}({\tilde r_e})} \right],
\label{eq_conform}
\eea
where $D_{KL}$ is, by definition, positive. This is a measure of how the non-equilibrium distribution $p({\tilde r_e})$ is different from the equilibrium distribution $p_{\rm eq}({\tilde r_e})$. With increasing $\Pe$ from zero, $D_{KL}$ initially increases to reach a maximum in the SAS state near $\Pe \approx 1$. This denotes the most non-equilibrium state, with the most compact spirals (smallest $\la \tilde r_e^2\ra$) spinning with the highest persistence (smallest $D_r^s$). At higher $\Pe$, as the spirals open up, $D_{KL}$  decreases. However, inertia suppresses the difference from equilibrium -- with increasing inertia, the magnitude of $D_{KL}$ even in the SAS state near $\Pe \approx 1$ decreases. This is another example of inertial equilibration found in other contexts in recent studies of active matter~\cite{Khali2024, Mandal2019, Patel2023, patel2024exact}.

\subsubsection{Radius of gyration tensor}

\begin{figure}[t!]
\includegraphics[width=\columnwidth,height=!]{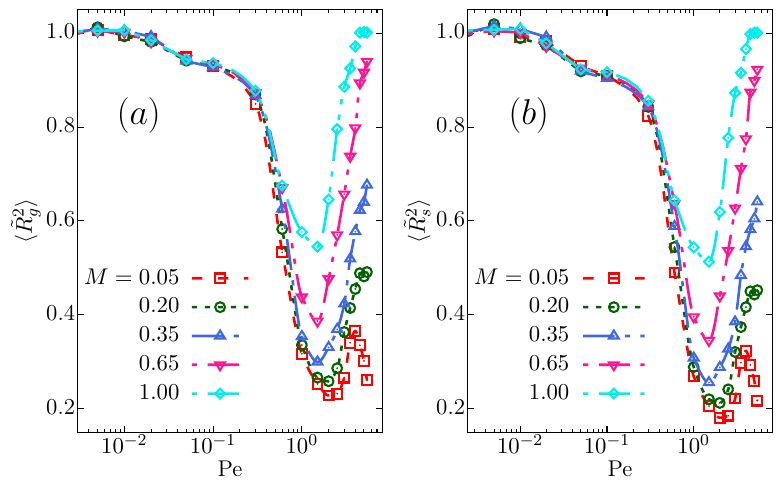}
\caption{(colour onlie) Scaled radius-of-gyration, $\la \tilde R_g^2 \ra$, and shape anisotropy, $\la \tilde R_s^2 \ra$ as functions of $\Pe$ for different $M$ values mentioned in the legend.}
\label{fig_Rg_Rs}
\end{figure}

Here, we consider the radius of gyration tensor to analyze the polymer size and shape further. This is given by
\bea
S = \f{1}{N} \begin{pmatrix}
\sum_i (x_i - x_{\rm cm})^2 & \sum_i (x_i - x_{\rm cm}) (y_i - y_{\rm cm}) \\\\
\sum_i (x_i - x_{\rm cm}) (y_i - y_{\rm cm}) & \sum_i (y_i - y_{\rm cm})^2
\end{pmatrix},
\label{eq_Rg_tensor}
\nn\\
\eea
where $\rv_i=(x_i,y_i)$ denotes the instantaneous position of $i$-th monomer and $\rv_{cm}=(x_{\rm cm},y_{\rm cm})$ denotes the position of the polymer center of mass $\rv_{cm}= (1/N) \sum_i \rv_i$. The larger and smaller eigenvalues of the matrix are $\l^+$ and $\l^-$, respectively. They can be used to determine the polymer size $R_g^2 = \l^+  + \l^-$ and shape anisotropy $R_s^2 = \l^+ - \l^-$,  with $R_s^2 = 0$ denoting a circular shape. 
In Fig.~\ref{fig_Rg_Rs} we show the variation of scaled mean quantities $\la \tilde R_g^2 \ra = \la R_g^2 \ra/\la R_g^2 \ra_{\Pe=0}$, and $\la \tilde R_s^2 \ra = \la R_s^2 \ra/\la R_s^2 \ra_{\Pe=0}$, with $\Pe$ for different $M$. Both size and shape show a non-monotonic, associated with the re-entrant transition to spiral and open chain. Lower inertia stabilizes the non-equilibrium spiral structures better, which is why, with increasing $M$, the minimum value of both the size and shape anisotropy increases.

\section{Conclusion}
\label{sec_discuss}

In summary, we presented a detailed phase diagram identifying the influence of inertia in the activity-induced transition to a spinning achiral spiral (SAS) state of a semi-flexible, self-avoiding chain. Inertia destabilizes SAS towards a motile open chain (MOC) at higher activity. We presented analytical arguments identifying the motile open chain (MOC) to SAS transition and the SAS to MOC re-entrant transition. This second transition is entirely an inertial effect and was absent in the overdamped limit. The dynamics at different states were analyzed using the mean-squared displacement of the center of mass and the evolution of the mean local tangents. We identified the SAS state to be the most out-of-equilibrium state using a Kullback-Leibler divergence measure from the equilibrium end-to-end distribution.  As our calculations showed, with increasing inertia, the SAS gets more and more destabilized, and the difference from equilibrium conformations decreases. It will be interesting to determine entropy production to characterize the notion of departure from equilibrium, further~\cite{Fodor2016, Chaudhuri2014, Ganguly2013}. Our predictions are amenable to direct experimental verification using macroscopic active chains, e.g., chains of hex-bug nano~\cite{Zheng2023}.

\section{Acknowledgement}

CK thanks Rajneesh Kumar for helping with the figures. The numerical simulations were performed using SAMKHYA, the High-Performance Computing Facility provided by the Institute of Physics, Bhubaneswar. DC acknowledges a research grant from the Department of Atomic Energy (1603/2/2020/IoP/R\&D-II/150288) and thanks the International Centre for Theoretical Sciences~(ICTS-TIFR), Bangalore, for an Associateship.

\appendix

\section{Mean turning number for spiral state}
\label{app_spiral}
\begin{figure}[htbp] 
\includegraphics[width=\columnwidth,height=!]{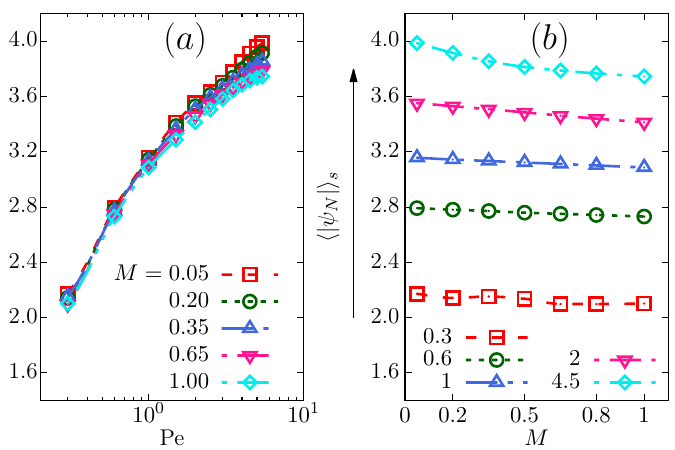}
\caption{(colour onlie) Mean absolute turning number for spiral state, $\la | \ord | \ra_s$, $(a)$ as function of $\Pe$ for different $M$ values, and $(b)$ as function of $M$ for different $\Pe$ values are plotted with corresponding legend.}
\label{fig_app1}
\end{figure}
We calculate the mean absolute numerical values of the turning number, $\ord$, for spiral state only, which denotes the degree of winding of the spiral structure as functions of $\Pe$ and $M$. 
As Fig.~\ref{fig_app1}$(a)$ shows, the average magnitude of turning number $\la |\ord| \ra$ increases as a function of $\Pe$ for every $M$, which means higher activity increases the turning number, creating a more compact structure of spirals. 
In Fig.~\ref{fig_app1}$(b)$, we observe that  $\la |\ord| \ra$ decreases with inertia at a given $\Pe$ because of the swelling due to inertial recoils. The relative change is more prominent at higher $\Pe$ values.

\section{Estimates of spiral state probability}
\label{app_prob}
\begin{figure}[htbp] 
\includegraphics[width=8cm]{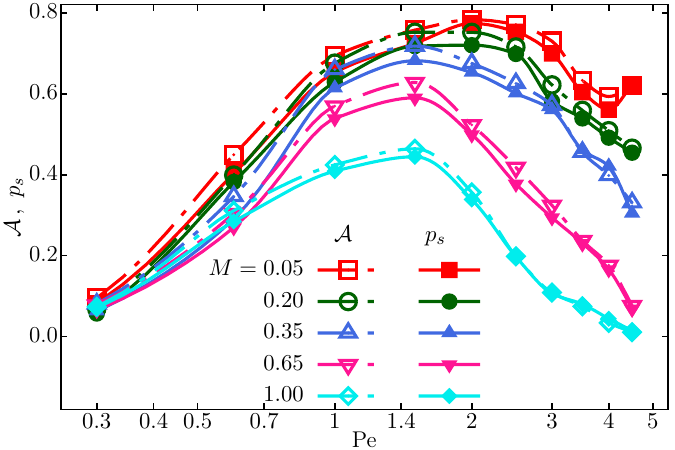}
\caption{(color online) Comparison between the probability of spiral state extracted from fitting the numerical data with Eq.\eqref{eq_triple_gauss}, $\pa$, and from the fitting of two-time correlation of Eq.\eqref{eq_corr_fitting}, $p_s$, as a function of $\Pe$ for different $M$ values. Legends in the plot denote the corresponding $M$.}
\label{fig_app2}
\end{figure}
In this section, we compare the two independent estimates of the spiral probabilities, $\pa$ and $p_s$, extracted from two different measurements mentioned in the main text: $(1)$ fitting of Eq.\eqref{eq_triple_gauss} to the numerically calculated distributions of $\ord$, and $(2)$ fitting of Eq.\eqref{eq_corr_fitting} with numerically calculated two-time tangent correlation. They follow each other closely~(Fig.~\ref{fig_app2}), thereby indicating the two measures are of indeed the same quantity, the probability of the spiral state. This vindicates our approximate theoretical description developed in the main text. 

\newpage

\bibliographystyle{unsrt}

\end{document}